\documentclass[12pt]{article} 
\usepackage{epsfig} 
 
\def\beq{\begin{equation}}
\def\eeq#1{\label{#1}\end{equation}}
\def\eeqn{\end{equation}}

\def\beqa{\begin{eqnarray}}
\def\eeqa#1{\label{#1}\end{eqnarray}}
\def\eeqan{\end{eqnarray}}

\let\bar=\overbar

\def\etal{{\it et al.}}

\def\Dslash{\not{\hbox{\kern-4pt $D$}}}
\def\dslash{\not{\hbox{\kern-2pt $\del$}}}

\def\msb{{\bar{\ssstyle M \kern -1pt S}}}

\def\Title#1{\begin{center} {\Large {\bf #1} } \end{center}} 
 
\begin{document}

\Title{Polarization curves of GRB afterglows predicted by the universal 
 jet structure   } 
 
\bigskip\bigskip 
 
 
\begin{raggedright}   
 
{\it Elena Rossi $^1$ \index{Rossi, E.}, Davide Lazzati$^1$\index{Lazzati, D.},  
Jay D. Salmonson$^2$ \index{Salmonson Jay D.}  and  Gabriele Ghisellini$^3$ \\ 
$^1$Institute of Astronomy Madingley Road, Cambridge CB3 OHA, UK\\ 
$^2$Lawerence Livermore Laboratory, L-095, P.O. Box 808, Livermore, CA, 94551 \\  
$^3$Osservatorio Astronomico di Brera. Via E. Bianchi 46, I--23807 Merate, Italia \\} 
 
\bigskip\bigskip 
 
\end{raggedright} 
 
\section{Introduction} 
 
\begin{figure}[htb] 
\begin{center} 
\epsfig{file=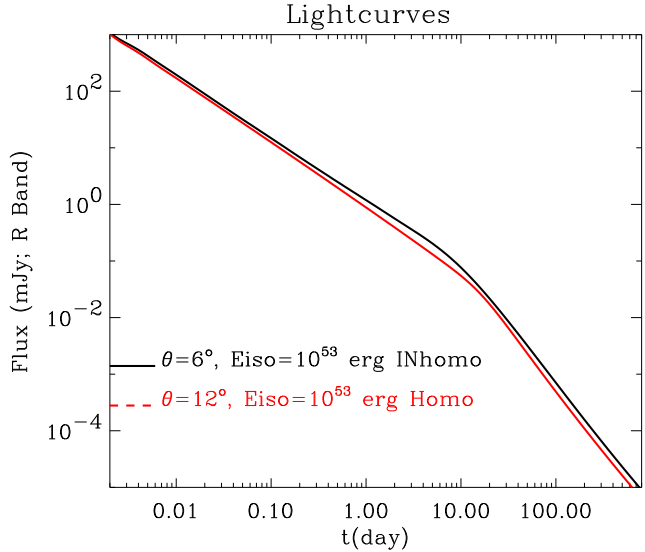,height=3in} 
\caption{Lightcurves for a structured jet 
(black line) seen at $6^{\circ}$ with $E_{iso}(6^{\circ})=10^{53} erg$ 
and an homogeneous jet (red line) seen on axis with an opening angle 
of $\theta_{jet}=12^{\circ}$ and isotropic equivalent energy 
$E_{iso}=10^{53}$ erg. The 2 lightcurves have the same break time and 
asymptotic slopes; they differ only up to a factor of 1.5 around the 
break time.} 
\label{fig1} 
\end{center} 
\end{figure}

Detections of linear polarization in the optical afterglow of GRBs are 
accumulating fast since the first measurements in GRB 990510 
~\cite{covino},~\cite{w99}.   
It was soon proposed that the observed polarization could arise 
from observing a collimated fireball with a slightly off 
axis line of sight (\cite{gl99},~\cite{sari99}). 
As a consequence, it was also realized that the degree  
of polarization could be connected with the achromatic break in the  
lightcurve expected in jetted fireballs. 
The polarization starts to grow when the observer perceives the nearest 
edge of the jet and the lightcurve begins to steepen; it has two 
maxima with the position angle being orthogonal to each other and then 
it slowly fades away while the jet front becomes entirely visible and 
the flux accomplishes the transition between the two asymptotic 
power--laws (Fig.4 in~\cite{gl99}).  Recently a different structure for the 
emission pattern has been proposed: a jet with a brighter (maybe 
faster) spine surrounded by dimmer (and slower) wings, with a standard 
energy reservoir~\cite{ppl},~\cite{rlr02},~\cite{jay},~\cite{zm},. 
The main parameters of such a structured jet are the angular size of 
the core $\theta_{\rm core}$ and the wings luminosity 
distribution. We~\cite{rlr02} have shown that if  the 
luminosity goes as $\theta^{-2}$  
 for $\theta>\theta_{\rm core}$ 
the lightcurves from such a jet are 
virtually indistinguishable from an homogeneous one (see ~\ref{fig1}) and 
both models can reproduce the break time--luminosity 
correlation~\cite{frail},~\cite{pk01}. Actually while in the standard 
model the time at which there is a steepening in the power--law decay 
of the afterglow, is related to the cone angle of the jet, in the 
structured jet such a break in the afterglow lightcurve occurs at a 
time that depends on the viewing angle. Instead of implying a range of 
intrinsically different jets -- some very narrow, and others with 
similar power spread over a wider cone -- the data on afterglow breaks 
are consistent with a standardized jet, viewed from different 
angles.  Since each luminosity is univocally related to a viewing 
angle, it is possible to calculate the predicted luminosity function 
~\cite{rlr02}; in the simplest version of this model, $n(L)\propto L^{-2}$.  
This prediction, if robustly confirmed by observations, could support the universal jet structure,  
but it cannot rule out the standard model. 
 
For this reason and the similarity of the lightcurves it is important 
to look for observations that can discriminate between the two 
different structures and we show in the following that 
polarization can be such a powerful tool. 
 
\section{The model} 
 
 We assume that the jets emerging from the central engine of GRBs 
are characterized by the following distributions in energy and initial Lorentz factor: 
\begin{equation} 
E_{iso}=E_{c}/[1+ \left(\theta/\theta_{c}\right) ^{2}] 
\label{eq:E} 
\end{equation} 
\begin{equation} 
\Gamma_o=\Gamma_{c}/[1+ \left(\theta/\theta_{c}\right) ^{\alpha_{\Gamma}}]. 
\label{eq:G} 
\end{equation} 
For simplicity we assume azimuthal symmetry.  We take $\frac {\Delta
log(\Gamma_o)}{\Delta \theta}\le 1$, therefore the regions with
different $\Gamma_o$ and energy are causally disconnected and they
evolve independently.  We can therefore treat separately the evolution
of each point of the jet, assuming adiabatic expansion.  Our code
integrates numerically the equation of relativistic energy
conservation and calculates consistently the evolution of the jet
aperture, if sideway expansion is considered (for the homogeneous jet
see e.g.~\cite{r99},~\cite{kp00}).  If and how post shock pressure
gradients develop in the case of a structured jet is a many parameters
problem, that can be solved only through hydrodynamic simulations.  We
therefore use a parametric analytic treatment of the sideway
expansion, where we can choose how the lateral velocity change with
$\theta$ and its maximum value.  We calculate the comoving frame
intensity assuming the standard synchrotron equations
(~\cite{pk00}~\cite{gs02}).  To compute the polarization vector we
assume a magnetic field configuration which corresponds to the
compression in one direction of an initially tangled magnetic field: 
it has some degree of alignment seen edge on while it is still completely  
tangled on small scales in the uncompressed plane.  
The maximum value of polarization $P0$ is 
carried by the light coming from an angle of $1/\Gamma$ with the line 
of sight.  We take $P0=60\%$, that corresponds to a completely ordered 
magnetic field in the sky plane.  
In order to compute lightcurves and polarization curves, 
local luminosities and polarization vectors are then summed over  
equal arrival time ($T$) surfaces,  
\begin{equation} 
 T=t_{\rm lab}-{r\over c}  \cos(\tilde{\theta}), 
\label{eq:sup} 
\end{equation} 
where $r$ is the radial distance from the source,  
$\tilde{\theta}$ is the angular distance from the line of sight and  
 \begin{equation} 
t_{\rm lab}=\int \frac{dr}{\beta c},    
\end{equation} 
is the time in the laboratory frame.

In this proceeding we show only the results for a non lateral
expanding jet, evolving in a constant density medium.  A more complete
treatment (with sideway expansion) will be completed soon ~\cite{r02}
 
\begin{figure}[htb] 
\begin{center} 
\epsfig{file=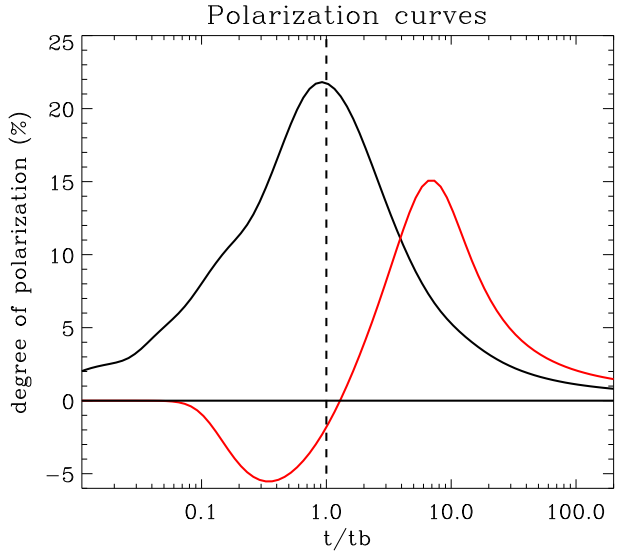,height=3in} 
\caption{  
Polarization curves for a structured jet (black line) and an
homogeneous jet (red line) seen at $\theta=0.67 \theta_{jet}$ (the
average off-axis angle).  The parameters are the same as Fig.1.  The
maximum value is $P0=60\%$, that corresponds to a completely ordered
magnetic field in the sky plane.  These curves can therefore be
considered as upper limits.  For $P>0$ the polarization vector lies on
the plane containing the line of sight and the jet axis; for $P<0$ it
is rotated by $90^\circ$. The most noticeable differences between the
structured and the homogeneous jet are that in the former case there
is not change in the polarization angle and the maximum is reached
around the break time.}
\label{fig2} 
\end{center} 
\end{figure} 

\section{Results} 
 
\begin{figure}[htb] 
\begin{center} 
\epsfig{file=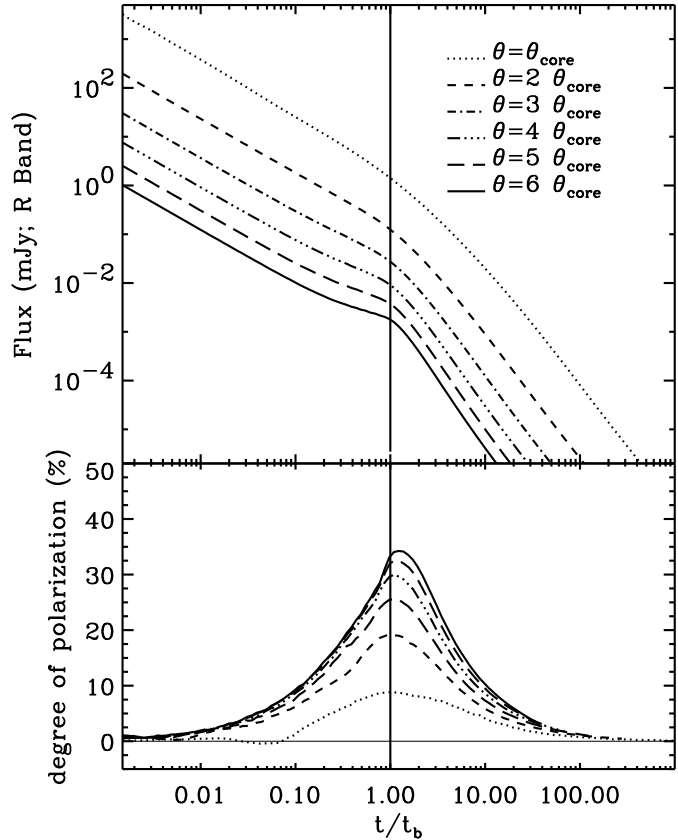,height=4in} 
\vspace{.5in} 
\caption{Structured jet with parameters: 
 $E_{core}=10^{54}$ erg, $\theta_{out}=30^{\circ}$, 
 $\theta_{c}=3^{\circ}$, $\Gamma_c=10^{4}$, $\alpha_{\Gamma}=2$, 
 $\epsilon_{e}=0.1,\epsilon_{B}=0.01, n=1$ cm$^{-3}$. See text for discussion.} 
\label{fig3} 
\end{center} 
\end{figure}

The results are summarized in Fig.~\ref{fig3}, 
where we show the lightcurves  
for the total and the polarized fluxes
for different viewing angles. 
The crucial parameters is the ratio 
between the viewing angle and the size of the core: 
\begin{equation} 
\theta/\theta_{c}\simeq (t_{b}/t_{bc})^{1/2}, 
\end{equation} 
where $t_{bc}$ is the  achromatic break time  
expected in the afterglow lightcurve for on--axis observers.

Fig. 3 shows that the more   
detailed calculations we have now performed  
confirm all the main features described in our previous paper ~\cite{rlr02} 
and allow a more precise description of them. 
For $\theta/\theta_{core} \ge 4$ the ligthcurves 
exhibit a mild flattening around the break time, increasing 
with the off axis angle, while for $\theta / 
\theta_{core} < 4$ the flattening is not evident 
and  the curves show sharp breaks and they look very similar.  
This flattening is due to the light coming from the core of the jet.                       
This contribution reaches its maximum when $1/\Gamma_c\simeq \theta_o$,        
shortly after $1/\Gamma(\theta_o)\simeq \theta_o$ (see also Eq.8 in             
~\cite{rlr02}); morover its peak flux is comparable to the                     
line of sight contribution at that time.                                        
Therefore this excess modifies the total lightcurve                            
around the time break, and the result is a flattening.

For a fixed viewing angle the direction of the vector is constant, and 
the magnitude is characterized by one maximum around the break time, 
when the central and most luminous part of the jet becomes 
visible. While the rising and fading slopes of the curve are 
independent of the off--axis angle, the value of the maximum increases 
with the viewing angle (see lower panel of Fig.~\ref{fig3}).  This 
behavior is very different from the one predicted in the homogenous jet 
model, (see Fig.~\ref{fig2}), despite the similarity in the 
lightcurves (see Fig.~\ref{fig1}).  For this reason we suggest that 
a monitoring of the change in the degree of polarization within one 
burst, especially before and after the break, can help to 
discriminate between the two models. 
 
\bigskip 

\def\Discussion{ 
\setlength{\parskip}{0.3cm}\setlength{\parindent}{0.0cm} 
     \bigskip\bigskip      {\Large {\bf Discussion}} \bigskip} 
\def\speaker#1{{\bf #1:}\ } 
\def\endDiscussion{} 
 
\Discussion 
 
\speaker{J. Rhoads} You have taken an angular dependence of  
 $\theta^-2$ and a standard jet                
energy in your model. Are these taken for theoretical elegance, or are 
they demanded by the data? I.e., if we suppose exponents from 1.8 to 
2.2 and a factor of 3 or 10 spread in energy, is that consistent with 
the data?   
 
\speaker{E. Rossi} To be consistent with data the exponent can range  
between 1.5 and 2.2 (1 $\sigma$ error). (See Fig. 4 in ~\cite{rlr02}). The 
value of 2 corresponds to no spread in the observered $\gamma$-ray 
energy.

\speaker{C. Fendt}   
From astrophysical jet simulations are known quite well the profiles 
of the dynamical parameters accross the jet. These may, however, 
completely be disturbed in the shock. Therefore, from observations of 
the inhomogeneous shell we can hardly derive any clue on the jet 
formation in the central engine. Could you please comment on that? 
 
\speaker{E. Rossi} 
 The inhomogeneous model we propose holds from a radius of $\sim10^{13}$ cm 
 and as all the other models for the fireball evolution, the jet's 
 structure does not retain the imprint of the central object. From 
 such a radius ahead, nevertheless, the inhomogenities in the jet are 
 not destroyed by the shock because any patchesis causally 
 disconnected. 
\endDiscussion 
  

\begin{thebibliography}{99} 
 

\bibitem{covino} 
S. Covino, D. Lazzati, G. Ghisellini, \etal , 1999, A\&A, {\bf 348}, L1 

\bibitem{frail} 
 D.~A. Frail, \etal, 2001, ApJ,  {\bf 562}, L55 
\bibitem{gl99} 
G. Ghisellini, D. Lazzati, 1999, MNRAS,  {\bf 309}, L7 
\bibitem{gs02} 
J. Granot, R. Sari, 2002, ApJ,  {\bf 568}, 820 
\bibitem{kp00} 
P. Kumar, A. Panaitescu, 2000, ApJ,  {\bf 541}, L9 
\bibitem{m00} 
N. Masetti, \etal, 2000, A\&A, {\bf354}, 473 
\bibitem{pk00} 
A. Panaitescu, P. Kumar, 2000, ApJ, {\bf 543}, 66 
\bibitem{pk01}  
A. Panaitescu, P. Kumar, 2001,ApJ,{\bf560}, 49  
\bibitem{ppl} 
K. A. Postnov,  M. E.Prokhorov, V. M. Lipunov, 2001, Astronomy Report, {\bf45}, 236 
\bibitem{r99}]  
 J.~E. Rhoads, 1999, ApJ, {\bf 525}, 737 
\bibitem{rlr02} 
E. Rossi, D. Lazzati, \& J. M. Rees, 2002, MNRAS, {\bf332}, 945 
\bibitem{r02} 
E. Rossi, D. Lazzati, J.Salmonson \& G. Ghisellini, in prep. 
\bibitem{jay} 
J. D. Salmonson,2001, Apj, {\bf487}, L1 
\bibitem{sari99} 
R. Sari, 1999, ApJ,  {\bf524}, L43 
\bibitem{w99} 
  R. A. M. J. Wijers \etal., 1999, ApJ, {\bf523}, 33L
\bibitem{zm} 
 Zhang, B; Mészáros, P, 2002, ApJ, {\bf571}, 876Z 
 
 
\end{thebibliography}
\end{document}